\documentclass[prb,reprint,letterpaper,twocolumn,showpacs,amsmath,amssymb,aps]{revtex4-2}

\usepackage{braket}
\usepackage{graphicx}
\usepackage{dcolumn}
\usepackage{algorithm}
\usepackage{algpseudocode}
\algrenewcommand{\algorithmicrequire}{\textbf{Input:}}
\algrenewcommand{\algorithmicensure}{\textbf{Output:}}
\usepackage{bm}
\newlength{\figwidth}
\usepackage{hyperref}
\hypersetup{
   breaklinks=true,   
   colorlinks=true,   
   pdfusetitle=true,  
}
\begin{document}

\title{Simplified algorithms for adaptive experiment design in parameter estimation}
\author{Robert D. McMichael}
\email{rmcmichael@nist.gov}
\author{Sean M. Blakley}

\affiliation{National Institute of Standards and Technology, Gaithersburg, Maryland 20899, USA}

\date{\today}

\begin{abstract}
In experiments to estimate parameters of a parametric model, Bayesian experiment design allows measurement settings to be chosen based on utility, which is the predicted improvement of parameter distributions due to modeled measurement results.  In this paper we compare information-theory-based utility with three alternative utility algorithms. Tests of these utility alternatives in simulated adaptive measurements demonstrate large improvements in computational speed with slight impacts on measurement efficiency.
\end{abstract}

\maketitle


\section{Introduction}

Bayesian experiment design is a powerful method for adaptive measurements to estimate the parameters of a nonlinear model function. Measurements to estimate parameters are commonplace in the physical sciences, where the traditional approach is to automate measurements using a sequence of pre-selected settings (a static design), followed by least-squares fitting of a model function to the data. This measure-then-fit approach is simple and effective, and entirely appropriate for many applications, but when measurement resources are limited, efficiency becomes more important.

As a more efficient alternative to static designs, sequential experiment designs use accumulated data to inform measurement setting decisions. Whether by choosing optimal designs, or simply by avoiding wasteful efforts, sequential design generally requires fewer measurements to achieve a given level of precision. However, sequential design adds two statistical tasks that add computational cost to the measurement cycle. Each cycle or {\em epoch} includes a setting-decision/design task and a data analysis task in addition to setting adjustment and data collection. 

Optimal Bayesian experiment design has roots in information theory and decision theory\cite{lindley_measure_1956} and is well-suited for nonlinear model functions. Chaloner and Verdinelli\cite{chaloner_bayesian_1995} published a review of early work that has  become a touchstone in the field, and in recent decades, the availability of computational power has sparked a resurgence of interest in Bayesian optimal design. In ref.\ \onlinecite{overstall_approach_2018}, Overstall et al. provide a brief review of design methods, and computational methods have been reviewed by Ryan et al.\cite{ryan_review_2016}. More recently, publication of software has provided direct access to Bayesian design methods methods, potentially reducing the effort required for implementation.\cite{overstall_acebayes_2020,olofsson_gpdoemd_2019, mcmichael_optbayesexpt_2021,ha_phoenics_2018,granade_qinfer_2017}

The Bayesian optimal design method centers around locating maxima of the {\em utility}, $U(d)$, which expresses the goals of the measurement as a function of candidate setting designs $d$. Several authors have identified computation of the utility as a particularly difficult part of Bayesian experiment design that prohibits its use.\cite{overstall_approach_2018,ryan_review_2016,huan_simulation-based_2013,long_fast_2013} Importantly, for a sequential design to be preferable to a static design, the cost of implementing and running the design processes must not exceed the value of saved resources. For practitioners, utility calculations must be easy and fast.

In this paper, we test the performance of several utility algorithms on two physics-related parameter estimation problems. The utility algorithms range in complexity from a numerical evaluation of the information entropy change to a minimal working example. Surprisingly, we find that these methods achieve similar results on a per-measurement basis, while computation time varies widely. The simplest utility algorithm tested here is orders of magnitude faster than the information-theoretic version with only a slight decrease in measurement efficiency.

Section \ref{sec:background} provides background, including an overview of the sequential design approach, Bayesian inference methods and the information-theoretic utility function. Section \ref{sec:algorithms} presents the numerical utility algorithms used in testing, with results presented in \ref{sec:results}, followed by discussion in section \ref{sec:discussion}.

\section{Background\label{sec:background}}

In this section, we provide an overview of the statistical methods. Bayesian experiment design provides a design method based in decision theory,\cite{lindley_measure_1956} and is described in a review by Chaloner and Verdinelli. \cite{chaloner_bayesian_1995}. The measurement run comprises a sequence of measurement cycles, or {\rm epochs}, each including a design step to select measurement settings, a measurement using those settings and yielding new data, and analysis to incorporate the new data. The approach is simulation-based Bayesian experiment design\cite{muller_optimal_2004}
follows the examples provided by Granade et al.\cite{granade_robust_2012} and Huan and Marzouk\cite{huan_simulation-based_2013}.
Below, we describe the data analysis process followed by the design process, which is the main focus of the paper.

The design and inference processes rely on a measurement model
\begin{equation}
    y = f(\theta, d) + \eta,
    \label{eq:model_def}
\end{equation}
where $y$ is the measurement output, $f(\theta, d)$ is a deterministic, nonlinear function of parameters $\theta$ and settings $d$ modeling the mean measurement value. The model function $f(\theta, d)$ may be nonlinear and non-monotonic, but is assumed to be a well-behaved and accurate model of measurement results. Random variable $\eta$ has a zero-mean distribution $P_\eta(\eta)$ and models measurement noise. The form of (\ref{eq:model_def}) accommodates intrinsically probabilistic models, such as quantum mechanics.  Finally, $\theta$ is a vector of model parameters $\theta = \left\{\theta_1, \theta_2, \ldots\right\}$ treated as random variables with joint distribution $P(\theta)$. The object of the measurement run is to provide precise estimates of the parameters, or equivalently, to narrow $P(\theta)$ around the true parameter values.

The analysis in each design-measure-analyze epoch uses Bayesian inference to calculate the influence of new data on the parameter distribution. After $n$ measurements, we wish to infer the posterior parameter distribution $P_n(\theta) \equiv P(\theta | {\bm y}_n, {\bm d}_n)$ given accumulated measurement data ${\bm y}_n \equiv \{y_1, y_2, \ldots y_n\}$ and corresponding settings ${\bm d}_n \equiv \{d_1, d_2, \ldots d_n\}$. Bold font indicates accumulated data, while non-bold indicates data from a single epoch. The inference can be performed iteratively using Bayes rule,
\begin{equation}
    P(\theta | {\bm y}_n, {\bm d}_n) \propto
    P(y_n | \theta, d_n)\, P(\theta | {\bm y}_{n-1}, {\bm d}_{n-1})
    \label{eq:bayes_iter}
\end{equation}
beginning with prior parameter distribution $P_0(\theta)$.  The likelihood $P(y_n | \theta, d_n)$, of obtaining result $y_n$ is the same as the probability that the noise value satisfies (\ref{eq:model_def}), or
\begin{equation}
    \label{eq:likelihood_def}
    P(y_n | \theta, d_n) = P_\eta(y_n - f(\theta, d_n)).
\end{equation}
In this paper we will assume that measurement noise is normally distributed with standard deviation $\sigma_\eta$. The likelihood of result $y_n$ is then given by
\begin{equation}
    P(y_n | \theta, d_n) = \frac{1}{\sqrt{2\pi}\sigma_\eta}
    e^{[y_n - f(\theta, d_n)]^2 / 2\sigma_\eta^2},
    \label{eq:likelihood}
\end{equation}
providing a quantitative answer to the question ``How well do the parameters explain the data?''

The distribution of parameters $P(\theta)$ encapsulates the state of belief regarding different parameter values. For computational work, methods known variously as particle filters, swarm filters or sequential Monte Carlo methods have gained prominence as representations of the probability distribution.\cite{gordon_novel_1993, arulampalam_tutorial_2002, carpenter_improved_1999, elfring_particle_2021} For a model with $D$ parameters, the distribution is represented by $N_{\rm p}$ ``particles'' each with a vector $\theta$ corresponding to a point in $D$-dimensional parameter space and a corresponding weight $w$ such that the distribution is approximated by
\begin{equation}
    P(\theta) \approx \sum_{k=1}^{N_{\rm p}}\delta(\theta-\theta_k)w_k
\end{equation} 
with $\sum_i w_i = 1$. Here $\delta(\cdot)$ is the Dirac delta function.
In inference calculations, unnormalized weights $W_i$ are generated by
\begin{equation}
    W_{i,n} = w_{i,n-1} P(y_n | \theta_i, d_n) 
\end{equation}
yielding new weights
\begin{equation}
    w_{i,n} = W_{i,n} / \sum_i W_{i,n}.
\end{equation}

Resampling is a necessary maintenance task for particle filters that effectively reassigns computer memory from low-probability regions of parameter space to high-probability regions. Resampling is initiated whenever the effective number of particles,
\begin{equation}
    N_{\rm eff} = 1 / \sum w_i^2,
\end{equation}
falls lower than a threshold, typically set at $0.5 N_{\rm p}$. The covariance matrix ${\bm C}$ of the distribution is calculated for a later step, then $N_{\rm p}$ samples $\tilde\theta_j$ are drawn with replacement from the $\{\theta_i\}$ set with probability $w_i$. This selection process will tend to miss low-weight particles, while possibly choosing high-weight particles multiple times. The particles are then given small random displacements $\delta\theta$ to separate degenerate particles. The resampled set of particles $\theta_j$ are given by
\begin{equation}
    \theta_j = \tilde\theta_j + \delta\theta_j
\end{equation}
where $\delta\theta_j$ is a sample from a multivariate normal distribution with covariance matrix $\alpha{\bm C}$. We use $\alpha = 0.01$.

We turn next to the question of selecting a design $d_{n+1}$ for the next measurement from a finite number of setting combinations. The information-theoretic approach introduced by Lindley\cite{lindley_measure_1956} and followed by many others chooses information entropy as a measure of the parameter distribution quality and it follows that the predicted change in information entropy following a measurement is the utility, $U(d')$. The entropy ${\cal H}$ of a distribution $P(\cdot)$ is given by
\begin{equation}
    \label{eq:entropy_def}
    {\cal H}_x[P] = -\int  P(x) \log\left(P(x)\right)\,dx.
\end{equation}
The Kullback-Leibler divergence describing the change in information entropy in going from $P_{n}(\theta)$ to $P_{n+1}(\theta)$ is
\begin{equation}
    \label{eq:kld_def}
    KL = - \int P_{n+1}(\theta) \log\left[ \frac{P_{n}(\theta)}{P_{n+1}(\theta)}\right] d\theta,
\end{equation}
where $P_n(\theta)$ is the parameter distribution after $n$ epochs, and $P_{n+1}(\theta)$ is the projected distribution after getting a possible measurement result $y'$ using a candidate setting design $d'$. The ``prime'' notation indicates speculative values as opposed to values $d$ and $y$ which are measurement records.
Using Bayes rule, the resulting parameter distribution would be given by
\begin{eqnarray}
    P_{n+1}(\theta)  & \equiv & P(\theta | y', d', {\bm y}_n, {\bm d}_n) 
      \\
     & = & \frac{P(y' | \theta, d')}{P(y')} P(\theta | {\bm y}_n, {\bm d}_n )
      = \frac{P(y' | \theta, d')}{P(y')} P_n(\theta), \nonumber
\end{eqnarray}
and the Kullback-Leibler divergence would then be written as
\begin{equation}
    KL(d', y') = -\int \frac{P(y' | \theta, d')}{P(y')} P_n(\theta) 
    \log\left[\frac{P(y')}{P(y' | \theta, d')}\right] d\theta.
\end{equation}
Averaging over the possible $y'$ measurement values yields the utility $U(d')$
\begin{eqnarray}
    U(d') & = & -\int P(y'| d') \log\left[P(y'| d')\right]  dy'    \label{eq:utility_def} \\
    & & +
    \int P_n(\theta) \left\{\int P(y' | \theta, d') 
    \log\left[P(y' | \theta, d')\right]  dy'\right\} d\theta
    \nonumber 
\end{eqnarray}
with
\begin{equation}
    P(y'|d') = \int P(y' | \theta, d') P_n(\theta)\, d\theta.
    \label{eq:yprime}
\end{equation}
Equation (\ref{eq:utility_def}) is the conventional, information-theoretic utility of a candidate setting, and it is the starting point for the investigations in this paper. This is the utility expression we approximate and/or emulate with simplified expressions.

The utility  (\ref{eq:utility_def}) has an intuitive interpretation. In the second term of (\ref{eq:utility_def}) the distribution $P(y' | \theta, d')$ appearing in the bracketed integral is the forecast distribution of measured values given a set of parameters $\theta$ and design $d'$. But with $\theta$ and $d'$ fixed, the arguments of the model function are all fixed and the distribution of $y'$ values is the noise distribution, offset by the model function value. Therefore, the second term may be interpreted as the (negative) entropy of the measurement noise, averaged over parameters.

The first term in (\ref{eq:utility_def}) is also an entropy of the predicted measurement value distribution, but here the distribution includes noise and also the effects of the parameter distribution as written explicitly in (\ref{eq:yprime}). In total, the utility is the entropy of possible measurement outputs produced by parameter uncertainty and noise, discounted by the entropy of noise. 

The intuitive interpretation of (\ref{eq:utility_def}) is simply that useful settings are those where the spread or dispersion of the parameter distribution produces large variations in measurement predictions relative to the measurement noise.  In the following section we present utility variations that all share this interpretation, but differ in their metrics for dispersion.

\section{Methods\label{sec:methods}}

\subsection{Utility Expressions\label{sec:algorithms}}
This subsection describes utility algorithms based on, or at least inspired by, (\ref{eq:utility_def}), with the goal of practical application in laboratory settings. To be useful, design algorithms must produce benefits of measurement efficiency that outweigh the cost of calculation. To meet that goal, we place a heavy emphasis on simplicity and speed. 

In pursuit of simplicity and speed, we also permit ourselves to invent trial utility functions without rigorous derivation. To justify this approach, we argue that the penalty for less-than-perfect design choices is a mild decrease in measurement efficiency, without invalidating the measurement results. Acceptable measurements are frequently made using completely arbitrary settings, after all. So, while we relax requirements for rigor in the formulation of utility algorithms, we are careful to monitor precision and accuracy of the parameter estimation results in the results presented below.

Qualitatively, the utility given by (\ref{eq:utility_def}) suggests that the best measurements will be made where the dispersion of the parameter distribution has a large effect on measurement outcomes relative to the dispersion of measurement noise. In (\ref{eq:utility_def}), the dispersion measure is information entropy, made so by  the conventional choice of information entropy to gauge improvements in $P(\theta)$. With the algorithms below, we explore alternative dispersion measures in utility functions.

\subsubsection{KLD utility algorithm}

This calculation attempts to approximate Eq.\ (\ref{eq:utility_def}). For the first term, $N_{\rm s}$ samples are drawn from the prior $P_n(\theta)$ and from the noise distribution $P(\eta)$ to simulate measurement outcomes using the measurement model, (\ref{eq:model_def}). The simulated outcomes are samples from $P(y'|\theta, d')$. The differential entropy is estimated from these samples using either Ebrahimi's method\cite{ebrahimi_two_1994} or Vasicek's method\cite{vasicek_test_1976}. For the examples in this paper, the noise is normally distributed with standard deviation $\sigma_\eta$, independent of $\theta$ and $d$, so the second term in (\ref{eq:utility_def}) can be determined from the known properties of the noise distribution: ${\cal H}_\eta = \frac{1}{2}\log(2\pi e \sigma_\eta^2)$. 

Pseudocode for $U^{\rm KLD}(d')$ is provided in Algorithm \ref{alg:kld} of the appendix.

\subsubsection{Variance algorithm}

For the variance algorithm, we choose the logarithm of variance as a measure of dispersion\cite{solonen_simulation-based_2012}, yielding the utility 
\begin{equation}
    U^{\rm var}(d') = \frac{1}{2}\log\left[v_\theta(d') + v_\eta\right] - \frac{1}{2}\log(v_\eta),
    \label{eq:var_entropy}
\end{equation}
which simplifies to
\begin{equation}
     U^{\rm var}(d') = \frac{1}{2}\log\left[1 + \frac{v_\theta(d')}{v_\eta}\right].
\end{equation}
Here, $v_\eta$ is the variance of the noise and $v_\theta(d')$ is the variance of a distribution of noise-free model values, $P(y'|d', \eta=0)$.  Samples of this distribution are provided by the model function $f(\theta_i|d')$ using parameter samples $\{\theta_1 \ldots \theta_{N_s}\}$. The first term in (\ref{eq:var_entropy}) represents the combined effects of parameters and noise, and the second term isolates the contribution due to noise alone. The variance utility also corresponds to the Kullback-Leibler utility (\ref{eq:utility_def}) in the special case where all distributions are normal. For a normal distribution with standard deviation $\sigma$, the entropy is
\begin{equation}
    {\cal H} = \frac{1}{2}\log(2\pi e \sigma^2),
    \label{eq:normalentropy}
\end{equation}
where $\log(e) = 1$.

A feature of the variance algorithm is that the variances of the parameter distribution and noise distributions are simply additive, so that effects of measurement noise and effects of parameter dispersion are separated. In contrast, the contributions to entropy are not easily distinguished. Another advantage is that variance computes faster than entropy.  

Pseudocode for $U^{\rm var}(d)$ is provided in Algorithm \ref{alg:var} of the appendix.

\subsubsection{Pseudo-utility algorithm}

Recognizing that standard deviation is not always a good proxy for entropy, the pseudo-utility attempts to recapture some of the properties of entropy as a dispersion measure while preserving the separation of parameter effects and noise effects provided by variance. With the same form as the variance utility, the pseudo-utility is defined as
\begin{equation}
    U^{\rm psu}(d') = \frac{1}{2}\log\left[1 +\frac{v_{\cal H}(d')}{v_\eta}\right],
\end{equation}
where $v_{\cal H}$ is an effective variance derived from the entropy of the distribution of noise-free model values. As in the variance algorithm, model values  are generated by the model function $f(\theta_i|d')$ for parameter samples $\{\theta_1 \ldots \theta_{N_s}\}$ and $\eta=0.$ The effective variance $v_{\cal H}$  requires the entropy\cite{ebrahimi_two_1994,vasicek_test_1976} of the noise-free model value distribution ${\cal H}_{y'}(d', \eta=0)$ which is then then transformed into a variance, using (\ref{eq:normalentropy}), which yields
\begin{equation}
    v_{\cal H} = (2 \pi e)^{-1} \exp[2 {\cal H}_{y'}(d', \eta=0)].
\end{equation}
The pseudo-utility is pieced together without a strong claim for validity, but it has been used in previous work.\cite{mcmichael_sequential_2021}  

Pseudocode for $U^{\rm psu}(d)$ is provided in Algorithm \ref{alg:psu} of the appendix.

\subsubsection{Max-Min utility algorithm}

The max-min utility is designed for simplicity and fast computation using the range statistic as a measure of dispersion.  A relatively small set of  samples $\theta_j$, $\{\theta_1 \ldots \theta_{N_s}\}$ are drawn from $P(\theta)$ and a corresponding set of $N_s$ model values $\{y(d')\} \equiv \{y_1(d') \ldots y_{N_s}(d')\}$ is calculated using $y_j(d') = f(\theta_j, d')$. The range $t(d')$ is defined as the difference between maximum and minimum values in the set,
\begin{equation}
    t(d') = {\rm max}(\{y(d')\}) - {\rm min}(\{y(d')\}).
\end{equation}
Again following the form of the variance utility, the max-min utility is defined as
\begin{equation}
     U^{\rm mm}(d') = \frac{1}{2}\log\left[1 + \frac{t(d')^2}{v_\eta}\right].
\end{equation}

\subsubsection{Random design}
As its name suggests, the random design chooses randomly from among the candidate settings, uninfluenced by measurement data. Because the settings are chosen with uniform probability, the measurement results are expected to exhibit the same overall performance as repeated sweeps of the setting value.  However, the random approach avoids artifacts that might be generated by periodic repetition of the settings.

\subsection{Sample Reuse\label{sec:reuse}}

Improved methods have been proposed by several authors to compute the double integral in the utility expression (\ref{eq:utility_def}) with sufficient precision \cite{huan_simulation-based_2013,long_fast_2013,drovandi_improving_2018, beck_multilevel_2020,ryan_estimating_2003}. In order to serve its purpose in experiment design, an approximate utility function need only exhibit a maximum at at design that is close to the design that maximizes the true utility. The utility values are ultimately discarded, while the design determines the next measurement. To simplify calculation of (\ref{eq:utility_def}), Huan and Marzouk\cite{huan_simulation-based_2013} proposed using the same set of parameter samples both for calculation of (\ref{eq:yprime}) and for the averaging over parameters in (\ref{eq:utility_def}). Where possible, we take this approach one step further by using one set of random draws to calculate utility for all candidate settings. Some modifications would be needed if the noise distribution independent on $\theta$ or $d$. 

\begin{figure}
    \centering
    \includegraphics[width=\figwidth]{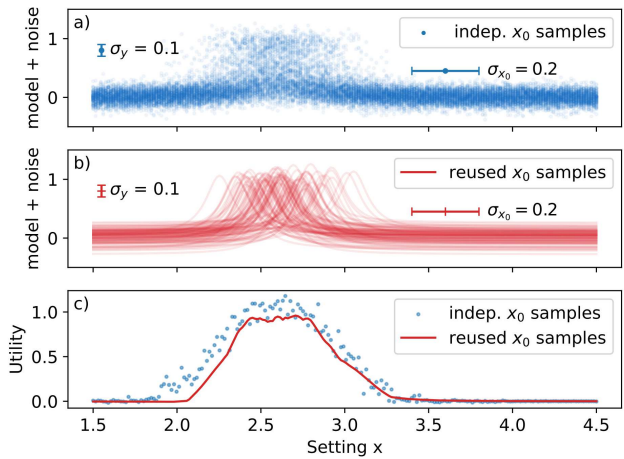}
    \caption{Illustration of the effects of parameter and noise sample reuse on utility function estimation. The model function is a Lorentzian given by (\ref{eq:reuse}). Panels a) and b) illustrate distributions of predicted measurement values $P(y'|x)$ using sets of $N_{\rm s} = 100$ samples each from normal distributions of center $x_0$ and added noise $\eta$. Standard deviations are indicated by bars. In a), a fresh set of samples is drawn for each value of setting $x$, while in b) a single set of samples is reused for all $x$. c) Corresponding estimates of utility, both yielding maxima near $x = 2.6$.  }
    \label{fig:reuseDraws}
\end{figure}

Fig.\ \ref{fig:reuseDraws} illustrates how sample reuse affects the utility function.  Panels a) and b) plot samples from $P(y'|x)$ for a Lorentzian model function
\begin{equation}
    y' = \frac{1}{[(x-x_0)/0.1]^2 + 1} + \eta
    \label{eq:reuse}
\end{equation}
with the peak center parameter $x_0$ and added noise $\eta$ both normally distributed. Panel a) plots simulated measurement outcomes where $N_s$ parameter samples and $N_s$ noise samples are drawn for each of $N_d$ candidate designs, requiring $2N_{\rm s}\cdot N_{\rm d}$ samples $[N_{\rm s} = 100; N_{\rm d} = 200]$. Fig.\ \ref{fig:reuseDraws}b) represents the same distribution, but single sets of $N_{\rm s}$ $x_0$ samples and $\eta$ samples are reused for all values of design variable $x$, requiring only $2N_s$ samples.  Fig.\ \ref{fig:reuseDraws}c) shows the effect of these contrasting approaches on the utility function using the KLD algorithm described below. Using new samples for each setting value ensures that the utility values are statistically independent, but introduces sampling noise that makes the maximum difficult to locate.\cite{ryan_review_2016} By reusing samples, the sampling noise is converted to a systematic error, and the point-to-point variation is suppressed. Both formulations will yield near-optimal setting values, but reusing samples greatly reduces computation time. Unless otherwise indicated we use $N_{\rm s} = 1000$ samples.

\section{Results and discussion\label{sec:results}}

\subsection{Lorentzian peak}

In this subsection, the simulated experiment is a Lorentzian absorption profile, with a single ``unknown'' peak-center parameter $\theta\equiv \{x_0\}$ and setting $d\equiv {x}$
\begin{equation}
    f_{\rm L}(x_0, x) = b + \frac{a}{[(x - x_0)/\Delta]^2 + 1}.
    \label{eq:lorentzian}
\end{equation}

Constants are background $b$, amplitude $a$, and half-width $\Delta$. The maximum Fisher information for the Lorentzian is found at $x = x_0 \pm \Delta / \sqrt{3}$, and the Cramer-Rao bound is given by
\begin{equation}
    \sigma_{x_0} \ge \frac{8}{3\sqrt{3}}\frac{\Delta}{a}\frac{\sigma_\eta}{\sqrt{n}}.
\label{eq:cramer-rao}    
\end{equation}

The Lorentzian function describes the frequency response of a damped harmonic oscillator, which appears frequently in many branches of science and engineering. Previous work demonstrated the effectiveness of sequential Bayesian experiment design on a similar system, estimating 5 parameters from measurements on triplets of similar peaks. Compared to a swept-setting approach, the Bayesian method required 40-fold fewer measurements to locate and measure to a comparable uncertainty.\cite{dushenko_sequential_2020} 

\begin{figure}[tb]
    \centering
    \includegraphics[width=\figwidth]{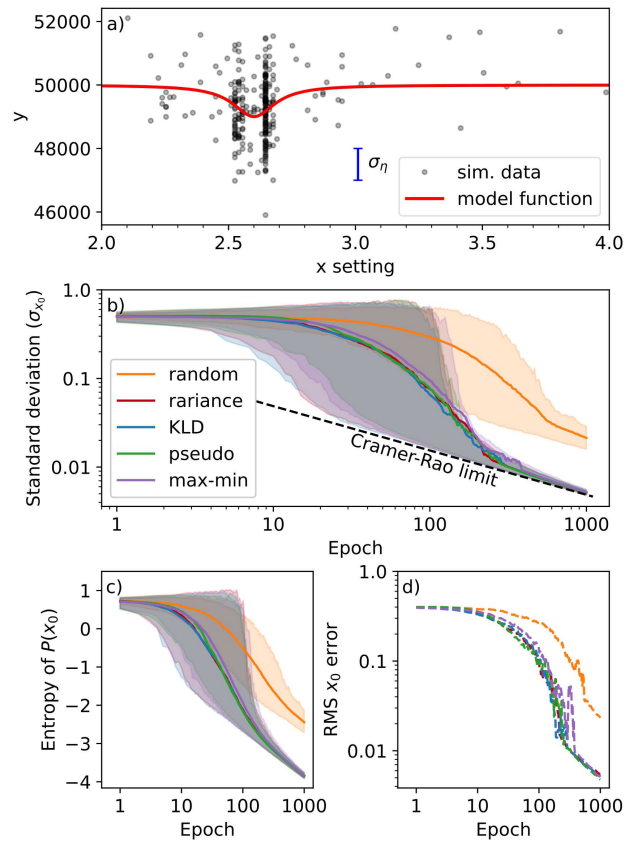}
    \caption{Simulated measurement runs to determine the center of a Lorentzian. a) The first 300 simulated data points of a selected run. The Lorentzian model function is plotted in red with the center parameter $x_0 = 2.6$, and the standard deviation of the noise is equal to the peak height. The adaptive design initially calls for settings over a broad range, but later repeats measurements on the sides of the peak after $\approx 100$ epochs. b) Standard deviation and c) entropy of $P(x_0)$ calculated using various utility algorithms. Mean (solid lines) and 5~\% to 95~\% credibility intervals (shaded areas) are calculated from 400 runs. The grey areas result from coincidence of the credibility intervals. d) Root-mean-square (RMS) error. Variance, KLD and pseudo algorithms use $N_s = 1000$ parameter samples; max-min uses $N_{\rm s} = 2$.}
    \label{fig:lotz_sigma}
\end{figure}

Fig.\ \ref{fig:lotz_sigma} presents results of simulated measurement runs using the sequential Bayesian design methods along with random setting selection for comparison.  The simulations use constants $b = 50000$, $a = -1000$, and $\Delta = 0.1$. Measurement data is simulated using true value $\theta^* = \{x_0 = 2.6\}$ and measurement noise is normally distributed with $\sigma_\eta = 1000$. The prior $P_0(x_0)$ is normal with mean 3.0 and standard deviation 0.5.
Fig.\ \ref{fig:lotz_sigma}a) shows data from a single run with the model ``true'' curve. After approximately 90 epochs in this run, the design algorithm focuses on settings close to the Fisher information maxima on the slopes of the peak/dip.

Panels b-d) of Fig.\ \ref{fig:lotz_sigma} support some of the main conclusions of this paper. In summaries of 400 measurement runs, panels b), c) and d) respectively trace the standard deviation and entropy of the parameter distribution, and the RMS error of the mean. 

A striking feature of Fig.\ \ref{fig:lotz_sigma} is that virtually identical results are obtained regardless of which utility algorithm is used. The sequential designs all follow similar approaches to the Cramer-Rao limit (\ref{eq:cramer-rao}) after a few hundred measurements. Compared to uninformed Random setting selection, all adaptive designs achieve $\sigma_{x_0}$ values that are a factor of 4 lower, corresponding to an order-of-magnitude savings in the number of measurements needed.

\begin{figure}[tb]
    \centering
    \includegraphics[width=\figwidth]{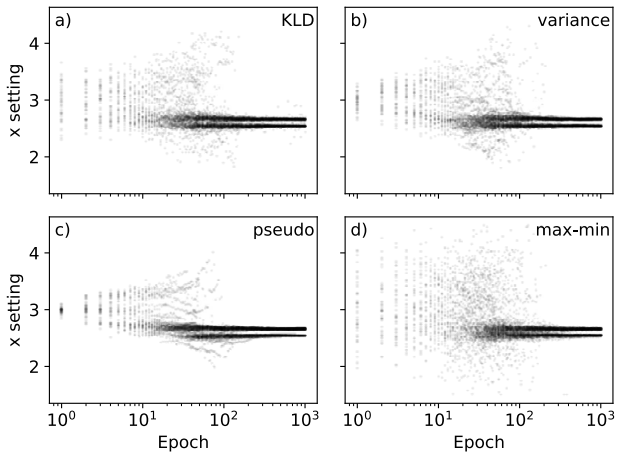}
    \caption{For each of the utility algorithms, setting values from 50 runs are superposed, demonstrating similar overall patterns in design choices. Epochs $> 50$, are sub-sampled for uniform appearance on the logarithmic scale.}
    \label{fig:lotz_xvals}
\end{figure}
Similarities in the the results are underscored by similarities in the design settings, as shown in Fig.\ \ref{fig:lotz_xvals}. Each run begins by forming a new particle filter using independent samples from the prior distribution, and each data simulation uses independent noise values, so different design choices are expected in different runs.  However, patterns in the design choices have strong similarities across the various utility algorithms, with an initial focus on $x$ values close to the center of the prior distribution at $x_0 = 3.0$, a broadening of the setting choices followed by rapid convergence to the settings for Fisher information maxima.

\begin{figure}[tb]
    \centering
    \includegraphics[width=\figwidth]{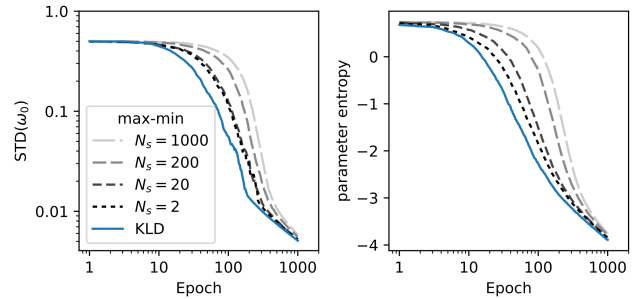}
    \caption{Tests of the max-min algorithm for varying numbers of parameter sample $N_{\rm s}$ including a) standard deviation and b) entropy of the $x_0$ parameter distribution. Performance approximates the KLD algorithm for small $N_{\rm s}$.}
    \label{fig:mm_ndraws}
\end{figure}
Figure\ \ref{fig:mm_ndraws} highlights behavior of the max-min utility algorithm and its dependence on the number of parameter samples $N_{\rm s}$ drawn for design calculations. The max-min algorithm uses the range of model function outputs over the parameter samples as a crude measure of dispersion due to parameter uncertainty. The results show that the minimum number of parameter samples ($N_{\rm s} = 2$) performs as well or better than larger sample numbers, and that its performance approaches that of the full KLD calculation.

\begin{figure}[b]
    \centering
    \includegraphics[width=\figwidth]{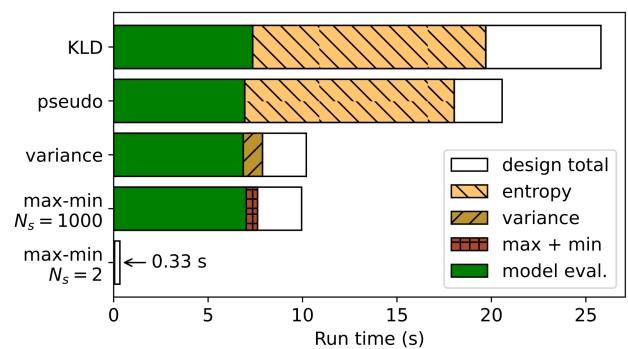}
    \caption{Comparison of design computation times among the utility algorithms in 1000-epoch runs. The runs simulate measurements to determine the center, $x_0$, of a Lorentzian peak using 200 candidate settings. Overall bar length indicates total design computation, with sub-tasks shown as segments. Model function and entropy calculations consume the majority of the total time. The max-min algorithm with $N_{\rm s} = 2$ executes very quickly.}
    \label{fig:lotz_speed}
\end{figure}
Figure \ref{fig:lotz_speed} shows the computation time required for the design algorithms and selected sub-tasks. Each of the  design calculations requires computation of the model function using $N_d = 200$ candidate designs for each of $N_{\rm s} = 1000$ parameter samples, followed by $N_d$ calculations of a statistic over $N_s$ samples.  The exception is the max-min algorithm using only $N_{\rm s} = 2$ parameter samples, which is unsurprisingly much faster. 

\subsection{Decaying sinusoid}

This subsection presents simulations of measurements to determine four parameters of a decaying sinusoid model function:

\begin{equation}
    f(\{h, c, \omega_0, T_2\}, \tau) = h + c\; \sin(\omega_0 \tau)\; e^{-(\tau/T_2)^2},
    \label{eq:Ramsey}
\end{equation}
where $\theta \equiv \{h, c, \omega_0, T_2\}$ are ``unknown'' parameters corresponding respectively to a background level, contrast, mean angular frequency, and dephasing with true values $h^* = 0.8, c^* = 0.13, \omega_0^* = 9.4$ and $T_2^* = 10$. The lone setting is $d \equiv \tau$. For each measurement, the system is initialized, the oscillations are started at $t=0$ and a single measurement is allowed after the a selected delay time $\tau$. For optimal measurements of $\omega_0$, the maximum slope $df/d\omega_0$ occurs for $\tau \approx T_2/\sqrt{2}$ and $\sin(\omega_0\tau) \approx 0$. For $n$ repeated measurements at this setting,
\begin{equation}
    \sigma_{\omega_0} = \frac{\sqrt{2e}}{c T_2}\frac{\sigma_y}{\sqrt{n}}.
    \label{eq:cr_ramsey}
\end{equation}
For optimal measurements of $T_2$, the maximum slope $df/dT_2$ occurs for $\tau \approx T_2$ and $\sin(\omega_0\tau)\approx \pm1$.

The decaying sinusoid describes a Ramsey measurement, which is the canonical method for measuring energy differences $\Delta E = \hbar \omega_0$ between quantum states\cite{degen_quantum_2017}, and it is closely related to the problem of measuring quantum phase. 

In (\ref{eq:cr_ramsey}), the frequency uncertainty is limited by measurement noise, but even a single noise-free measurement would yield ambiguous results since many values of $\omega_0$ can produce the same value of $y$. Interest in Ramsey measurements is driven by the possibility that the energy difference $\Delta E$ can be measured (i.e., disambiguated) with precision $\Delta E \propto 1 / \Delta t$ in a time $\Delta t$.~\cite{kitaev_quantum_1995,kitaev_quantum_1996} This behavior is termed Heisenberg scaling in analogy with the Heisenberg uncertainty principle, $\Delta E \Delta t \ge \hbar/2$. When readout fidelity is high, but not perfect, Heisenberg scaling can be approached using adaptive\cite{higgins_entanglement-free_2007,berry_how_2009,said_nanoscale_2011,cappellaro_spin-bath_2012,bonato_optimized_2016} and by optimized non-adaptive\cite{higgins_demonstrating_2009,nusran_high-dynamic-range_2012,waldherr_high-dynamic-range_2012} measurements. 

\begin{figure}
    \centering
    \includegraphics[width=\figwidth]{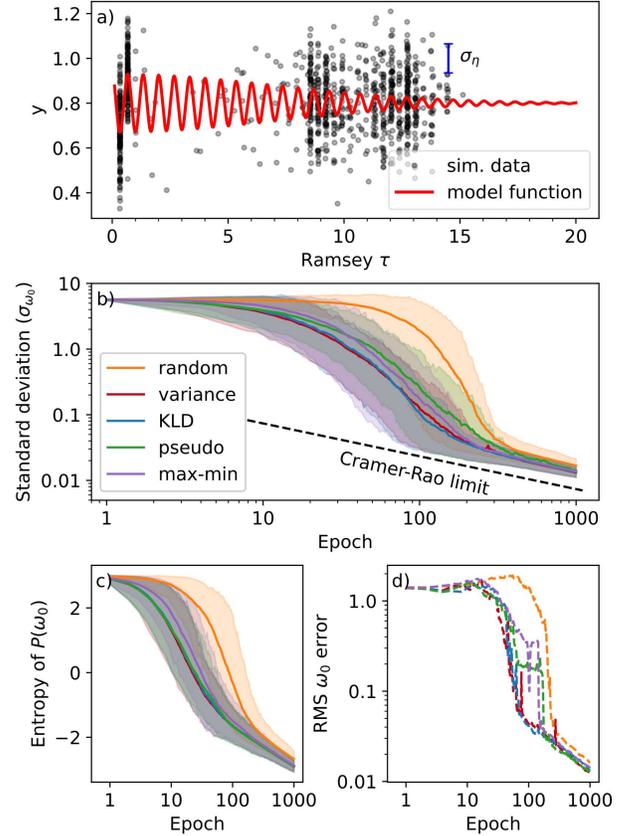}
    \caption{Simulated measurement run to determine four parameters of a decaying sinusoid. a) The model function is plotted in red as a function of setting $\tau$. The standard deviation of the noise is equal to the oscillation amplitude at $\tau = 0$. The adaptive design initially chooses low $\tau$ settings, but later favors $\tau \approx T_2$. Data from epochs later than 50 have been subsampled for clarity. b) Standard deviation and c) entropy of $P(\omega_0)$ of simulated measurements using various utility algorithms. Mean (solid lines) and 5~\% to 95~\% credibility intervals (shaded areas) are calculated from 400 runs. d) RMS error.}
    \label{fig:ramz_sigma}
\end{figure}
Figure \ref{fig:ramz_sigma} presents results of Ramsey sequence simulations.  Fig.\ \ref{fig:ramz_sigma}a) plots the ``true'' model function (\ref{eq:Ramsey}) and simulated data. Allowed settings cover the range $0.1 \le \tau \le 20$ in steps of 0.01. The simulated noise is normally distributed, with $\sigma_\eta = 0.13$. Fig.\ \ref{fig:ramz_sigma}b) compares the standard deviation of $\omega_0$ for various utility algorithms as a function of the number of measurement epochs. As with the results of the Lorentzian peak example, the utility algorithms achieve similar performance on a per-measurement basis. 

\begin{figure}
    \centering
    \includegraphics[width=\figwidth]{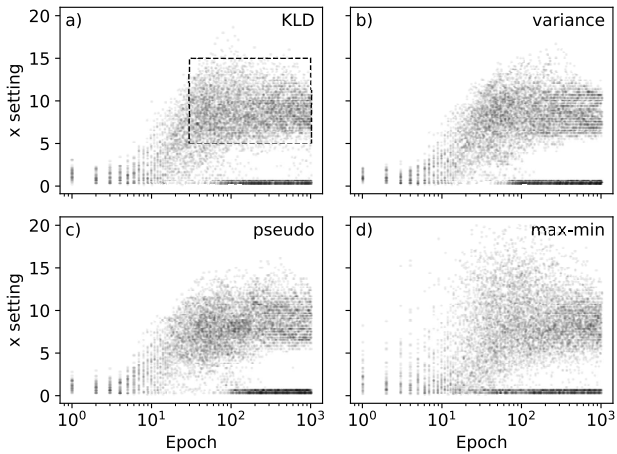}
    \caption{Setting values plotted for 50 runs each using the indicated utility algorithms. In later epochs, the algorithms choose small $\tau$ and also $\tau$ distributed across ranges, for example the range indicated by the dashed rectangle in panel a). The chosen $\tau$ values tend to fall at local extrema of the model function and local extrema of slope. These correlations are more clearly visible in Fig.\ \ref{fig:ramz_sigma}a)}
    \label{fig:ramz_xvals}
\end{figure}
The setting values generated by the algorithms are plotted in Fig.\ \ref{fig:ramz_xvals}. Strong similarities between the designs are visible, despite differences in the algorithms.  All the design algorithms begin with low-$\tau$ measurements and increase $\tau$ gradually.  After 100 epochs, the algorithms select concentrated measurements at low $\tau$ and distributed measurements in the region roughly $5\lesssim\tau\lesssim 15$, for example within the boxed region in Fig.\ \ref{fig:ramz_xvals}a).

Within the diffuse region of settings, there are strong correlations between $\tau$ values and the sinusoid phase. In Fig.\ \ref{fig:ramz_sigma}a), close examination shows that the low-$\tau$ settings are correlated with maxima and minima of the model function. For $8 \lesssim \tau \lesssim 11$, $\tau$ values are correlated with zero crossings, suggesting optimal settings for determining $\omega_0$. Also, $\tau$ above 11 are correlated with maxima and minima suggesting optimal settings for determining $T_2$.

In contrast to the Lorentzian model results in Fig.\ \ref{fig:lotz_sigma}, where there are large performance differences between the adaptive algorithms and random parameter selection, the random design applied to the Ramsey model achieves similar performance to adaptive designs after approximately 400 epochs. We attribute the relatively good performance of random setting selection here to the feature-rich structure of the Ramsey model function and the larger number of parameters. Taken together, these factors create a situation where virtually any selection of $\tau$ is somewhat informative. The broad distributions of settings chosen by adaptive designs also suggest that utility is not concentrated on a few settings in the decaying sinusoid measurement.

\begin{figure}
    \centering
    \includegraphics[width=\figwidth]{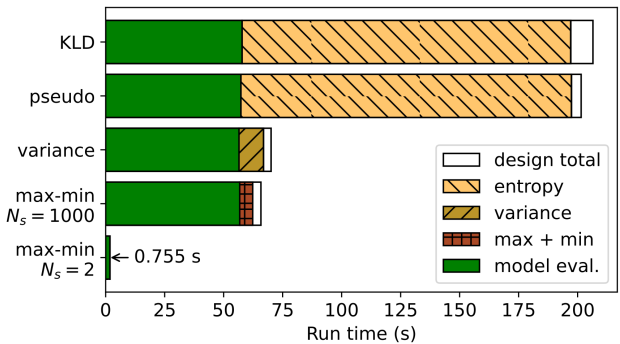}
    \caption{Comparison of design task computation among the utility algorithms in 1000-epoch runs using the decaying sinusoid model with four parameters and 1991 setting values. Within the total time, model function and entropy calculations are the most costly parts of the algorithms. The max-min algorithm with $N_{\rm s} = 2$ executes very quickly.}
    \label{fig:ramz_speed}
\end{figure}
Figure \ref{fig:ramz_speed} shows the computation time required for the design algorithms and selected sub-tasks. Each of the  design calculations requires computation of the model function using $N_d = 1991$ candidate designs for each of $N_{\rm s} = 1000$ parameter samples, followed by $N_d$ calculations of a statistic over $N_s$ samples.  The exception is the max-min algorithm using only $N_{\rm s} = 2$ parameter samples. 

\section{Discussion\label{sec:discussion}}

This paper compares results of simulated measurement runs using two nonlinear experiment models and four utility algorithms using different statistical functions, but all treating the utility as a measure of the dispersion of forecast measurement results relative to the dispersion of measurement noise. In the regime where signal contrast was on par with measurement noise, the five utility algorithms generated similar design choices and similar reductions in parameter variance, entropy, and RMS error. While the KLD utility has a solid theoretical basis, the KLD algorithm does not produce dramatically better designs than simplified algorithms.

We point to the max-min algorithm with $N_{\rm s} = 2$ as an important outcome of this study. This algorithm pushes the utility-as-dispersion idea to its simplest extreme, calculating the model function for each candidate design and only two parameter samples. This simplified utility algorithm calculates much faster than the other tested algorithms with only a slight decrease in efficiency. 

An interesting point is that the max-min algorithm shows poorer performance with larger numbers of parameter samples. Fig.\ \ref{fig:mm_ndraws} shows that using $N_{\rm s} > 2$ makes less efficient design decisions, despite more thorough sampling of parameters. We explain the relative inefficiency of $N_{\rm s} > 2$ by noting that the max-min algorithm disregards representative parameter values from the center of a distribution when $N_{\rm s} > 2$, and instead takes direction from the atypical extreme values.   

By reusing parameter samples in utility calculations, we make utility estimates correlated across designs. The benefits of this technique are first, that the utility estimates become smooth functions of the setting value with easily located maxima, and second, that fewer random parameter samples are required. Although these benefits come at the cost of a systematic error, the effects of the systematic error are transient, because new parameter samples are drawn for each epoch. The $N_{\rm s} = 2$ max-min algorithm especially benefits from suppressed sampling noise, because it lacks the natural noise suppression provided by large sample numbers. 

In this paper, we have demonstrated speed and quality of the max-min $N_{\rm s} = 2$ algorithm on CPU hardware. We speculate that this algorithm will be well suited for implementation on FPGAs or other dedicated hardware with fixed-point arithmetic, because the max-min statistic requires only subtraction and absolute value operations. However, practical implementation of a dedicated system would still face challenges in implementing a parameter probability density representation and mathematical operations to support model function and likelihood calculations.


\newpage
\appendix

\section{Pseudocode\label{app:pseudocode}}

\begin{figure}[h]
\begin{algorithm}[H]
\caption{Kullback-Leibler algorithm}\label{alg:kld}
\begin{algorithmic}
\Require{Parameter distribution $P(\theta)$}
\Require{Noise distribution $P(\eta)$}
\For{$j \gets 1 \ldots N_s$}
  \State{Sample $\theta_j$ from $P(\theta)$ parameter distribution}
  \State{Sample $\eta_j$ from $P(\eta)$ from noise distribution}
  \For{all candidate settings $d_i$}
    \State{Simulate measurement $y_{i,j} \gets {\rm model}(\theta_j, d_i) + \eta_j$}
    \State{}
    \Comment{Reusing $\theta_j$ and $\eta_j$}
  \EndFor
\EndFor
\For{all candidate settings $d_i$}
  \State{Estimate entropy from samples $h_{y,i} = {\cal H}(y_{i,1} \ldots y_{i,N_s})$}
  \State{Estimate entropy of noise distribution $h_{\eta,i} \gets {\cal H}[P(\eta)]$}
  \State{Calculate utility $U^{\rm KLD}_i \gets h_{y, i} - h_{\eta, i}$}
\EndFor
\State {Find max utility $i_{\rm best} \gets$ argmax$(U^{\rm KLD}_i)$}
\Ensure{corresponding setting $d_{i_{\rm best}}$}
\end{algorithmic}
\end{algorithm}
\end{figure}

\begin{figure}[h]
\begin{algorithm}[H]
\caption{Variance utility algorithm}\label{alg:var}
\begin{algorithmic}
\Require{Parameter distribution $P(\theta)$}
\For{$N_s$ parameter samples $\theta_j$ drawn from $P(\theta)$}
    \For{all candidate designs, $d_i$}
        \State{Evaluate model $y_{i,j} \gets f(\theta_j, d_i)$}
        \State{}
        \Comment{reusing $\theta_j$ samples}
    \EndFor
\EndFor
\For{all candidate designs $d_i$}
    \State{Variance $v_{\theta,i} \gets {\rm Var}(y_{i,1} \ldots y_{i,N_s})$
     over parameters}
    \State{Variance of noise $v_{\eta,i} \gets {\rm Var}[P(\eta)]$}
    \State{Calculate utility $U^{\rm var}_i \gets \log[1 + v_{\theta,i}/v_{\eta,i}]/2$}
\EndFor
\State{Find max utility $i_{\rm best} \gets {\rm Argmax}(U^{\rm var}_i)$}
\Ensure{corresponding setting $d_{i_{\rm best}}$}
\end{algorithmic}
\end{algorithm}
\end{figure}

\begin{figure}[H]
\begin{algorithm}[H]
\caption{Pseudo-utility algorithm}\label{alg:psu}
\begin{algorithmic}
\Require{Parameter distribution $P(\theta)$}
\For{$N_s$ parameter samples $\theta_j$ drawn from $P(\theta)$}
    \For{all candidate designs, $d_i$}
        \State{Evaluate model $y_{i,j} \gets f(\theta_j, d_i)$}
        \State{}
        \Comment{reusing $\theta_j$ samples}
    \EndFor
\EndFor
\For{all candidate designs $d_i$}
  \State{Estimate entropy from samples $h_{y,i} = {\cal H}(y_{i,1} \ldots y_{i,N_s})$}
  \State{Calculate effective variance $v_{\theta,i}^{\rm eff} \gets \exp[2 h_{y,i}] / (2\pi e)$}
  \State{Variance of noise distribution $v_{\eta,i} \gets {\rm Var}[P(\eta)]$}
  \State{Calculate utility $U^{\rm PU}_i \gets \log[1 + v_{\theta,i}^{\rm eff}/v_{\eta,i}]/2$}
\EndFor
\State{Find max utility $i_{\rm best} \gets {\rm Argmax}(U^{\rm PU}_i)$}
\Ensure{corresponding setting $d_{i_{\rm best}}$}
\end{algorithmic}
\end{algorithm}
\end{figure}

\begin{figure}[H]
\begin{algorithm}[H]
\caption{Max-min utility}\label{alg:m_m}
\begin{algorithmic}
\Require{Parameter distribution $P(\theta)$}
\For{$N_s$ parameter samples $\theta_j$ drawn from $P(\theta)$}
    \For{all candidate designs, $d_i$}
        \State{Evaluate model $y_{i,j} \gets f(\theta_j, d_i)$}
        \State{}
        \Comment{reusing $\theta_j$ samples}
    \EndFor
\EndFor
\For{all candidate designs $d_i$}
    \State{Find maximum model value $y_i^{\rm max} 
            = {\rm max}(y_{i,1} \ldots y_{i,N_s})$}
    \State{Find minimum model value $y_i^{\rm min} 
            = {\rm min}(y_{i,1} \ldots y_{i,N_s})$}
    \State{Calculate range $t_i \gets  y_i^{\rm max} -  y_i^{\rm min})$}
    \State{Variance of noise $v_{\eta,i} \gets {\rm Var}[P(\eta)]$}
    \State{Calculate utility $U^{\rm mm}_i \gets 
        \log[1 + t_i^2/v_{\eta,i}]/2$}
\EndFor
\State{Find max utility $i_{\rm best} \gets {\rm Argmax}(U^{\rm mm}_i)$}
\Ensure{corresponding setting $d_{i_{\rm best}}$}
\end{algorithmic}
\end{algorithm}
\end{figure}

\bibliographystyle{aipnum4-2}
\bibliography{Utility}

\end{document}